\documentclass[10pt, conference, letterpaper]{IEEEtran}
\usepackage{balance}
\usepackage{cite}
\usepackage{epsfig}
\usepackage{times}
\usepackage{graphicx}
\usepackage[figuresright]{rotating}
\usepackage{amssymb}
\usepackage{amsmath}
\usepackage{setspace}
\usepackage{rotating}
\usepackage{multirow}
\usepackage{wrapfig}
\usepackage{booktabs}
\usepackage{array}
\usepackage{comment}
\usepackage{pict2e}
\usepackage{epstopdf}
\usepackage{multirow}
\usepackage{hhline}
\usepackage{subcaption}
\usepackage[]{nohyperref}  
\usepackage{url}  

\usepackage{algorithmic}
\usepackage{algorithm}
\floatname{algorithm}{Algorithm} \textfloatsep 0.5cm

\usepackage{colortbl}

\newcommand{\be}{\begin{equation}}
\newcommand{\ee}{\end{equation}}

\begin{document}
\title{Energy-efficient Analog Sensing for Large-scale, High-density Persistent Wireless Monitoring}

\author{{Vidyasagar Sadhu, Xueyuan Zhao, and Dario Pompili}\\
Department of Electrical and Computer Engineering, Rutgers University--New Brunswick, NJ, USA\\
E-mails: xueyuan\_zhao@cac.rutgers.edu, vidyasagar.sadhu@rutgers.edu, pompili@ece.rutgers.edu\\
}


\maketitle

\thispagestyle{empty}

\begin{abstract}
The research challenge of current Wireless Sensor Networks~(WSNs) is to design energy-efficient, low-cost, high-accuracy, self-healing, and scalable systems for applications such as environmental monitoring. Traditional WSNs consist of low density, power-hungry digital motes that are expensive and cannot remain functional for long periods on a single charge. In order to address these challenges, a \textit{dumb-sensing and smart-processing} architecture that splits sensing and computation capabilities among tiers is proposed. Tier-1 consists of dumb sensors that only sense and transmit, while the nodes in Tier-2 do all the smart processing on Tier-1 sensor data. A low-power and low-cost solution for Tier-1 sensors has been proposed using Analog Joint Source Channel Coding~(AJSCC). An analog circuit that realizes the rectangular type of AJSCC has been proposed and realized on a Printed Circuit Board for feasibility analysis. 
A prototype consisting of three Tier-1 sensors (sensing temperature and humidity) communicating to a Tier-2 Cluster Head has been demonstrated to verify the proposed approach. Results show that our framework is indeed feasible to support large scale high density and persistent WSN deployment.
\end{abstract}


\section{Introduction}\label{sec:introduction}
\textbf{Overview: }
The research and engineering objective of continuous monitoring of the physical world through minuscule ``smart dust'' motes in the `90s helped spur nearly two decades of exciting research in Wireless Sensor Networks~(WSNs). Some of that research has been successfully commercialized, while some other has been a precursor to recent advances in the ``Internet of Things.'' Still, the vision of a large-scale WSN comprising tens of sensors per square meter while being robust to sensing/communication/computation failures remains far from a reality. Indeed, even though Hewlett-Packard's much touted \emph{Central Nervous System for the Earth} project hopes to deploy billions of sensors all over the planet~\cite{HP-CeNSE}, its first commercial partnership with Shell for seismic monitoring still relies on motes that require VHS-sized enclosures~\cite{HP-Shell}. The fundamental reason for this large gap between vision and reality of WSNs is that the \textit{design and production of motes combining sensing, communication, and computation capabilities into a single, miniature platform (the cornerstone of traditional WSN paradigm) that can remain operational for months, if not years, on a single charge, that can self heal from internal failures, and that are still cheap is an extremely difficult engineering challenge}.


\textbf{Motivation: }
%
Nowadays motes are composed of digital processors, multiple Analog-to-Digital Converters~(ADCs) and wireless transceivers. These digital motes sense the environment but also carry out digital
communications and computations, both of which also require high bit resolution for high precision and dynamic range. Digital motes as a result tend to be power hungry. On the other hand, sensing and basic communications can be carried out by power-efficient analog sensors. Moreover, the spatio-temporal characteristics of the underlying phenomenon being monitored by a Cyber Physical System~(CPS) are seldom, if ever, known in advance. In order to support low-cost, high-confidence, and scalable CPS's, therefore, it is desired that today's digital sensor motes adapt their temporal resolutions and bit precisions during the operation of the WSN. However, the reality is that state-of-the-art motes are ``monolithic" due to various cost and design considerations. Consequently, a careful and often irreversible choice of design parameters for digital motes is made prior to the WSN deployment, resulting in either \emph{over-} or \emph{under-provisioning}: the former leads to heavy under-utilization of motes, while the latter results in low sensing resolution and accuracy. Due to the cost of digital sensors, it would not be feasible to deploy thousands of such sensors to monitor the environment. Hence, these questions are raised: can we have a low-cost and low-power solution for the sensing, and meanwhile being able to compress the sensing source and perform coding to combat against the distortion in the wireless channel?

\textbf{Our Vision:} To address these questions, we propose a \emph{modularized sensing architecture} (Fig.~\ref{fig:big_picture}) that represents \emph{a paradigm shift} from the traditional two-tier WSN architecture (with monolithic sensing and computing digital motes reporting to a ``sink") to a three-tier architecture. In this architecture, while the ``sink'' tier---consisting of powerful fusion center(s) to perform central processing and higher control---still exists, we split the traditional ``digital motes'' tier into two tiers consisting of low-cost, energy-efficient, analog sensors at tier-1 (to support the sensing and communication functionalities) and resource-rich digital cluster head nodes at tier-2 (to support processing and control).
Our architecture uses a large number of low-cost/low-power/low-accuracy analog sensors ($\approx 130 \mu W \text{without radio power}, \$5$) instead of a small number of high-cost/power-hungry/high-accuracy digital motes.
The \textit{low-cost} factor enables us to deploy these sensors in \textit{large scale} and \textit{high-density} thereby providing high \textit{spatial accuracy}. The \textit{low-power} on the other hand means the sensors need not be put to sleep thereby providing high \textit{temporal accuracy} unlike the current digital nodes which go to sleep occasionally to conserve power. These vast number of analog sensors at tier-1 do only task of sensing and transmitting (which we call \textit{dumb-sensing}), while the onus of processing (which we call \textit{smart-processing}) this sensor data for extracting useful information lies on powerful digital nodes at tier-2. This paper mainly focuses on realizing this low-cost and low-power analog sensing at tier-1. For this, we design sensor nodes with Shannon-mapping capabilities. The Shannon mapping~\cite{Shannon49} is a low-complexity technique for Analog Joint Source-Channel Coding (AJSCC)~\cite{Hekland05}; it can compress two or more signals into one (introducing controlled distortion) while also staying resilient to wireless channel impairments. We have currently used Frequency Modulation~(FM), for the sensors to communicate to a digital cluster head in tier-2, due to its impressive performance under low SNR conditions.

\textbf{Our Contributions}
can be summarized as follows:
\begin{itemize}
  \item We propose a \textit{dumb-sensing} and \textit{smart-processing} framework for wireless sensor networks that splits sensing and computational tasks between energy-efficient low cost sensors (Tier-1) and powerful digital nodes (Tier-2) respectively. We focus mainly on Tier-1 analog sensing in this paper.
  \item We propose to use Analog Joint Source-Channel Coding~(AJSCC) for Tier 1 to realize low cost and low-power consumption;
  \item We verify the feasibility of our proposal through simulations and experiments using simple tier-1 prototype developed by us.
\end{itemize}

\textbf{Paper Outline: }
The remainder of this paper is structured as follows. In Sect.~\ref{sec:prop_arch}, we present our three-tier architecture for WSNs and discuss its features. In Sect.~\ref{sec:prop_ckt_top}, we discuss our solution to low-power and low-cost tier-1 sensing using AJSCC and its parameter optimization.
In Sect.~\ref{sec:perf_eval}, 
we discuss the hardware prototype developed and present some system-level results to study the feasibility of our proposal. Finally, in Sect.~\ref{sec:conc}, we conclude the paper and discuss our future work.


\section{Three-tier Low-power Sensing Architecture}\label{sec:prop_arch}
\begin{figure}
\begin{center}
\includegraphics[width=3.4in]{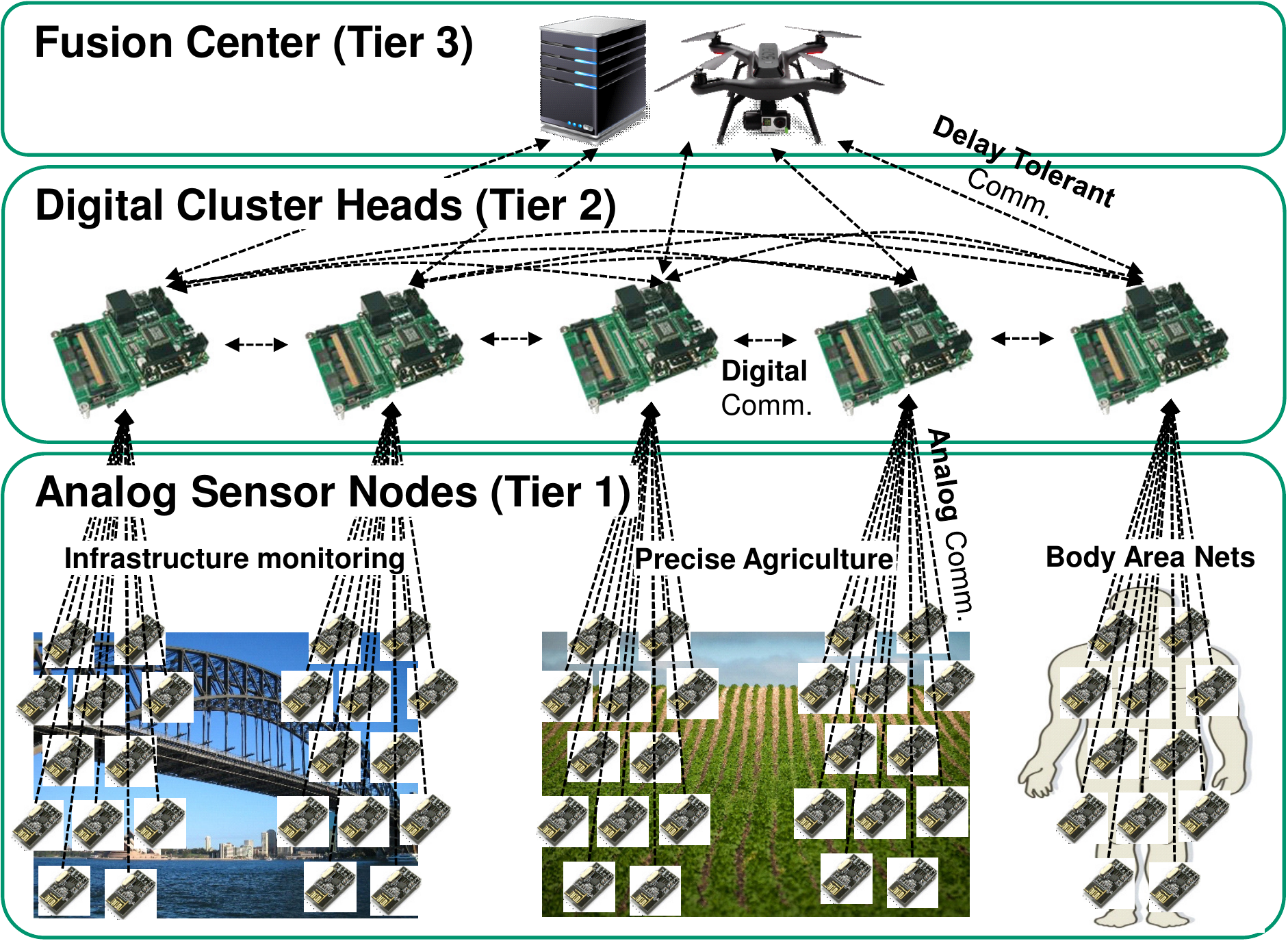}
\end{center}
\caption{\emph{Tier-1 analog sensor nodes} perform sensing and communicate with a digital Cluster Head (CH) via Analog Joint Source-Channel Coding. \emph{Tier-2 digital CHs} perform computation/processing and control, and digitally communicate among themselves and with the \emph{Tier-3 fusion center}.}\label{fig:big_picture}
\end{figure}

Our architecture breaks away from the past design goal of homogeneous WSNs comprising high-power, resource-rich digital motes with integrated sensing, communication, and computation
capabilities. Instead, we advocate a three-tier architecture (Fig.~\ref{fig:big_picture}) that corresponds
to a high density of extremely low-cost and low-power \emph{``dumb'' analog sensors} with limited communication capabilities in Tier-1 and a low density of \emph{``smart'' digital Cluster Heads (CHs)} with advanced communication and computation capabilities in Tier-2. Tier-3 consists of a fusion center (can be a server or a mobile drone) having similar functionalities of a Tier-2 sink in traditional WSN architecture. While the communication among digital CHs in tier-2 can be digital, that between tier-2 and tier-3 can be delay tolerant as the fusion center may not be always available. We would like to clarify that the general idea of tiered architectures for WSNs is not new to the research community; see, e.g., \cite{Ye.etal.Conf2002,Shah.etal.AHN2003,Wang.etal.EJAiSP2003}. The main idea of this paper lies in its use of dumb all-analog sensors for low-power sensing and communication at Tier 1 which should, in principle, enable large-scale, high-density wireless monitoring. 


Signals from multiple analog sensors will be multiplexed and detected at the digital CHs, which will be equipped with high-rate, high-resolution Analog to Digital Converters~(ADCs) and digital transceivers to communicate with neighboring CHs, and processors to run fault-detection/data-fusion algorithms (``Smart Processing''). Finally, data processed by the CHs will be retrieved through the fusion center (e.g., a mobile node such as a drone)---which may not be always connected to the network---that reconstructs the phenomenon of interest and also possibly generates control commands to support ``closed-loop scenarios''. Our architecture will help to \emph{simultaneously} improve sensing resolution (spatial and temporal), accuracy, and energy efficiency. Smart processing techniques at tier-2 consist of detecting faulty sensors, denoising, filtering, encryption (if needed), compression and other techniques needed to process ``big-data'' coming from tier-1 sensors. Developing these data processing/computational techniques for tier-2 CHs will be considered as future work and outside the scope of this paper.

Upgrades to the sensing tier (Tier~1, composed of analog sensors) can be made \emph{independently} without affecting the computing tier (Tier~2, composed of digital CHs), and vice-versa. This is possible because the sensors only sense and do not have any intelligence. So it should be easy to replace them with other sensors. Similarly, upgrades to the computing capabilities of the CHs can be made without affecting the analog sensors. Three-tier architecture also makes the WSNs \emph{incrementally} deployable: any WSN following this new paradigm can easily coexist (backward compatible) with already deployed WSNs in existing CPS's. In a world of incremental deployment, we believe this marks a major contribution towards low-cost, high-confidence, scalable CPS's. We claim that the broader applications of our proposed architecture include low-cost, high-confidence monitoring for urban infrastructure, precision agriculture, intelligent transportation systems, and military surveillance, to name a few.

In agreement with the Latin phase, \emph{Natura non facit saltus} (``nature does not make jumps"), our sensors are analog, as measurements are taken for the real world, which is inherently analog.
Compared to Commercial Off-The-Shelf (COTS) motes, the energy consumption of pure analog sensing can be much lower: an all-analog node should consume, in theory, on the order of $mW$ or less power (which is comparable to the power that can be harvested using, for example, a compact solar panel), while a COTS node’s power consumption is typically on the order of several tens of $mW$. Considering the sensors' low-power and low-cost benefits these can be either rechargeable or disposable.
Broader applications of our approach include low-cost, high-confidence monitoring for intelligent transportation systems, military surveillance, urban infrastructure, precision agriculture, and even body area networks (with no fusion center). We believe it will provide significant benefits in terms of ease of upgrade and scaling out of WSNs in addition to adaptive sensing resolution, accuracy, and energy efficiency. 



\section{Low-power Tier-1 Sensing}\label{sec:prop_ckt_top}
We first discuss the reasons for choosing AJSCC by giving an overview of the
potential advantages of AJSCC.
Then, we introduce our proposed circuitry for realizing AJSCC.

\begin{figure}
\begin{center}
\includegraphics[width=3.4in]{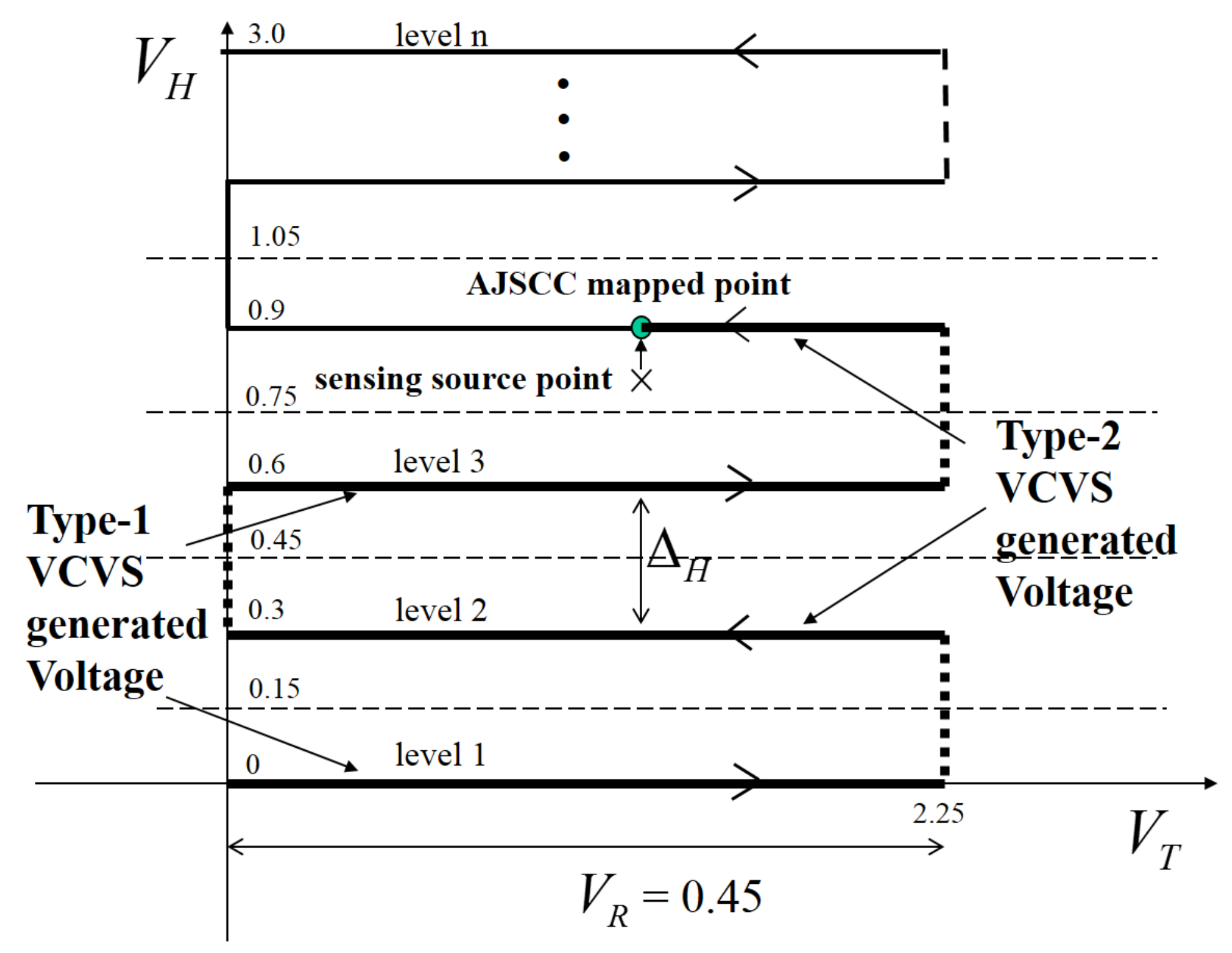}
\end{center}
\caption{Shannon's Rectangular Mapping. Sensed point is mapped to a point closest on the rectangular curve and the length of the curve from origin to the mapped point (bold part) is transmitted instead of two sensed values. Odd level voltages are generated using Type-1~VCVS while that of even level are generated using Type-2~VCVS.}\label{fig:rec_mapping}
\end{figure}

\begin{table*}
  \centering
  \caption{Relationship to prior work.}\label{table:compare}
  \vspace{-0.1in}
  \footnotesize
\resizebox{\textwidth}{!}{%
\begin{tabular}{|p{0.08\textwidth}|p{0.45\textwidth}|p{0.4\textwidth}|}
    \hline
    \textbf{Related work} & \textbf{Description} & \textbf{Comparison with the proposed research} \\ \hline\hline
    Patent~\cite{stopler14} & Digital video transmission by AJSCC/Shannon mapping & Digital implementation of AJSCC \\ \hline
    Patent~\cite{Park12} & Wireless analog sensors for implantation application & No source and channel coding considered \\ \hline\hline
    Paper~\cite{Garcia11} & Software-defined radio system for AJSCC in indoor channel & Digital implementation of AJSCC \\ \hline
    Paper~\cite{Romero14} & Digital optical communication system with AJSCC encoding for image transmission & Digital implementation of AJSCC \\ \hline\hline
    Product~\cite{wsn340}  & WSN340: Active MCU power consumption: 1.1~$\mathrm{mW}$ & \multirow{4}{*}{\scriptsize \parbox{0.4\textwidth}{Our proposed sensor will consume $\approx130~\mu{\mathrm{W}}$ with state-of-the-art low-power components (OpAmps, etc.)~\cite{Zhao16}. There is potential to reduce this even further ($<50~{\mu \mathrm{W}}$) when all the functionalities are integrated into a monolithic component using analog IC design.}} \\ \cline{1-2}
   Product~\cite{cosen} & Mantaro CoSeN: Active MCU power consumption: 2.4~$\mathrm{mW}$ & \\  \cline{1-2}
   Product~\cite{telosb} & Telos RevB: Active MCU power consumption: 6~$\mathrm{mW}$  &  \\  \cline{1-2}
   Product~\cite{mica2} & MICA2: Active MCU power consumption: 26.4~$\mathrm{mW}$ & \\ \cline{1-2}
    \hline
  \end{tabular}}
  \vspace{-0.2in}
\end{table*}

\subsection{Analog Joint Source Channel Coding~(AJSCC) in WSNs}\label{sec:motivation}
AJSCC adopts Shannon mapping as its encoding method~\cite{Fresnedo13,Hekland05}. Such mapping, in which the design of \emph{rectangular (parallel) lines} can be used for 2:1 compression
(Fig.~\ref{fig:rec_mapping}), was first introduced by Shannon in his seminal 1949 paper~\cite{Shannon49}. Later work has extended this mapping to a \emph{spiral type} as well as to N:1 mapping~\cite{Brante13}. The Shannon mapping has the two-fold property of (1) compressing the sources (by means of N:1 mapping) and (2) being robust to (wireless) channel distortions as the
noise only introduces errors along the parallel lines (or the spiral curve). Joint analog source-channel coding achieves optimal performance in rate-distortion ratio. It is known that to achieve optimal performance in communications using \emph{separate} source and channel coding, complex encoding/decoding and long-block length codes (which cause significant delays) are required.

It is worth noting that there are information-theoretic analyses on whether the separate source-channel coding deployed in real communication systems is optimal or not. Nazer and Gastpar~\cite{Nazer08} argue that, for a Gaussian sensor network, analog-scaled transmission performs exponentially better than a separate source-channel coding system.
In~\cite{Gastpar08}, a sensing system is studied for single memoryless Gaussian source, multiple independent sensors with Gaussian noise, and a cluster head node with standard Gaussian multiple-access channel. It is stated that the optimal communication strategy is analog scaled transmission, where each sensor transmits a scaled signal of the noisy sensed signal to the communication channel connecting sensor nodes with the cluster head node. Also, analog communications can be optimal in certain circumstances, e.g., when Gaussian samples are transmitted over an Additive White Gaussian Noise~(AWGN) channel and the source is matched to the channel. As mentioned earlier, AJSCC is resilient to channel noise because the channel noise only introduces errors along the spiral curve or the parallel lines. In contrast, linear mapping techniques such as Quadrature Amplitude Modulation~(QAM) have errors spread on the constellation plane. Therefore channel noise has less effects on the error performance for Shannon mapping than linear modulation schemes.

All these reasons motivated us to choose the \emph{combination} of analog communication \emph{and} Joint Source-Channel Coding~(JSCC), hence \emph{AJSCC}, as sensing/transmission scheme for our low-cost, low-power Tier-1 analog sensors. AJSCC performs analog compression at the symbol level. Also, the fact that symbols are memoryless makes it a low-latency and low-complexity solution that is very suitable for practical implementations. Last, but not least, AJSCC schemes can also achieve performance close to the Optimal Performance Theoretical Achievable~(OPTA) for Gaussian sources~\cite{opta1,opta2}, thus making it very attractive for our Tier-1 sensing.

AJSCC requires simple compression and coding, and low-complexity decoding. To compress the source signals (``sensing source point"), the point on the space-filling curve with minimum Euclidean distance from the source point is found (``AJSCC mapped point"), as in Fig.~\ref{fig:rec_mapping}. The two most-widely adopted mapping methods are rectangular (Fig.~\ref{fig:rec_mapping}) and spiral shaped: in the former, the transmitted signal is the ``accumulated length'' of the lines from the origin to the mapped point; while in the latter it is the ``angle'' that \emph{uniquely} identifies the mapped point on the spiral. At the receiver (a CH), the reverse mapping is performed on the received signal using Maximum Likelihood~(ML) decoding. The error introduced by the two mappings is controlled by the spacing $\Delta_H$ between lines and spacing
$\Delta_S$ between spiral arms, respectively: with smaller $\Delta_H$ (or $\Delta_S$), approximation noise is reduced; however, channel noise is increased as a little variation can push the received symbol to the next parallel line (or spiral arm). In addition, the mapping signal range would also increase, pushing more resources for transmission.

\subsection{Proposed Circuit Realization of AJSCC}\label{sec:prop_ckt}
A low-power, low-cost, and high-accuracy \emph{analog circuit} needs to be designed as existing AJSCC-hardware solutions are all digital and power hungry as they combine both sensing/communication and processing per the traditional architecture. For example, a Software-Defined Radio~(SDR) system to realize AJSCC mapping has been reported in~\cite{Garcia11}. The mapping was also recently implemented in an optical digital communication system in~\cite{Romero14} and has been combined with Compressive Sensing~(CS) in~\cite{Saleh12} to improve robustness against channel noise. Shannon mapping encoding was adopted in~\cite{stopler14} for a digital video transmission. All these design solutions use digital microcontrollers, which are quite power hungry: for example in~\cite{TI11}, with a $1.8~\mathrm{V}$ supply, the power consumption of a microcontroller alone can be as high as $450~\mathrm{mW}$ ($250~\mathrm{mA} \times 1.8~\mathrm{V}$); not to mention that the actual power consumed will be even greater when one considers other power-hungry digital components such as ADC/DAC/FPGA/DSP. Table~\ref{table:compare} compares our work with the existing patents, papers and products which are close to our work. To the best of our knowledge, none of them implemented Analog Joint Source Channel Coding using analog circuitry to achieve low-power, low-cost sensing as we did. While comparing our product with the TI sensor in~\cite{Romero14}, we understand that the TI sensor is doing many other processing jobs apart from just sensing, justifying it's cost and power consumption. However, it is to be noted that one of our main ideas in this paper is to deviate from such architecture, i.e., to split the two functionalities - sensing in tier-1 and processing in tier-2. We have also successfully demonstrated a working prototype using that architecture. Because of this, our sensors that are to be deployed on the field become less expensive and less power-hungry lending themselves well to high-density deployment which in turn leads to highly accurate spatial and temporal sensing of the environment.

Figure~\ref{fig:rec_mapping} shows the Shannon's rectangular mapping curve applied to the range of temperature voltage, $V_T$ and humidity voltage $V_H$. Let's denote the cross point as the actual sensed point. As Shannon Mapping maps the actual sensed point to the closest point on the rectangular curve, the circle denotes the mapped point. Hence 2D information consisting of $V_T$ and $V_H$ has been compressed to 1D information, the length of the curve from the origin to the mapped albeit with some quantization error. This length of the curve can henceforth be used for modulation and transmission instead of $V_T$ and $V_H$. The receiver upon receiving this 1D information decodes it back to 2D information using simple modulus calculation. The number of levels, the encoded length and the quantization error in $V_H$ are determined by resolution sought in $V_H$ ($\Delta_H$).

To the best of our knowledge, there has been no prior work on how to realize this length in practice using analog circuitry. We have come up with an innovative solution to calculate the length of this curve as a function of the mapped $V_T$ and $V_H$ values. Let's assume the mapped point lies on level 1. In this case the encoded length varies proportional to $V_T$. When the mapped point lies on level 2 it varies inversely proportional to $V_T$ (if the level 1 length is subtracted from it). In fact, this basic behavior can be observed at all odd and even levels respectively (i.e., assuming the total length of levels below it are subtracted from the encoded length). If we consider the mapped point lying on some arbitrary level, the encoded length would be equal to the sum of lengths of all levels below it and the length either proportional (odd level) or inversely proportional (even level) to $V_T$. This means there are two components that determine the encoded length - the level on which the mapped point lies and whether the level is odd or even. The latter is easily found if we assign odd and even indicators to all levels which is trivial when we know the number of levels (i.e., $\Delta_H$) in advance and we make this assumption as of now. The former can be found by comparing $V_H$ with threshold voltages of the levels (dashed lines in Fig.~\ref{fig:rec_mapping}). Hence \emph{each} level contributes one of these three values to the total encoded length: zero, partial (how much is based on whether odd or even level) and full level length. Using this idea we came up with a circuit realization for each level. For the partial length case to realize proportionality, we made use of a Voltage Controlled Voltage Source~(VCVS) which outputs voltage that is proportional to the input voltage. Let's call this Type-1~VCVS and for odd levels this can be used directly. However for even levels, we need another type of VCVS, we call it Type-2~VCVS that gives output inversely proportional to the input. For these reasons, we use odd (even) level and Type-1 (Type-2) level interchangeably. We sometimes refer the combination of two consecutive levels (Type-1 and Type-2) as a single stage.

\begin{figure}
\begin{center}
\includegraphics[width=3.4in]{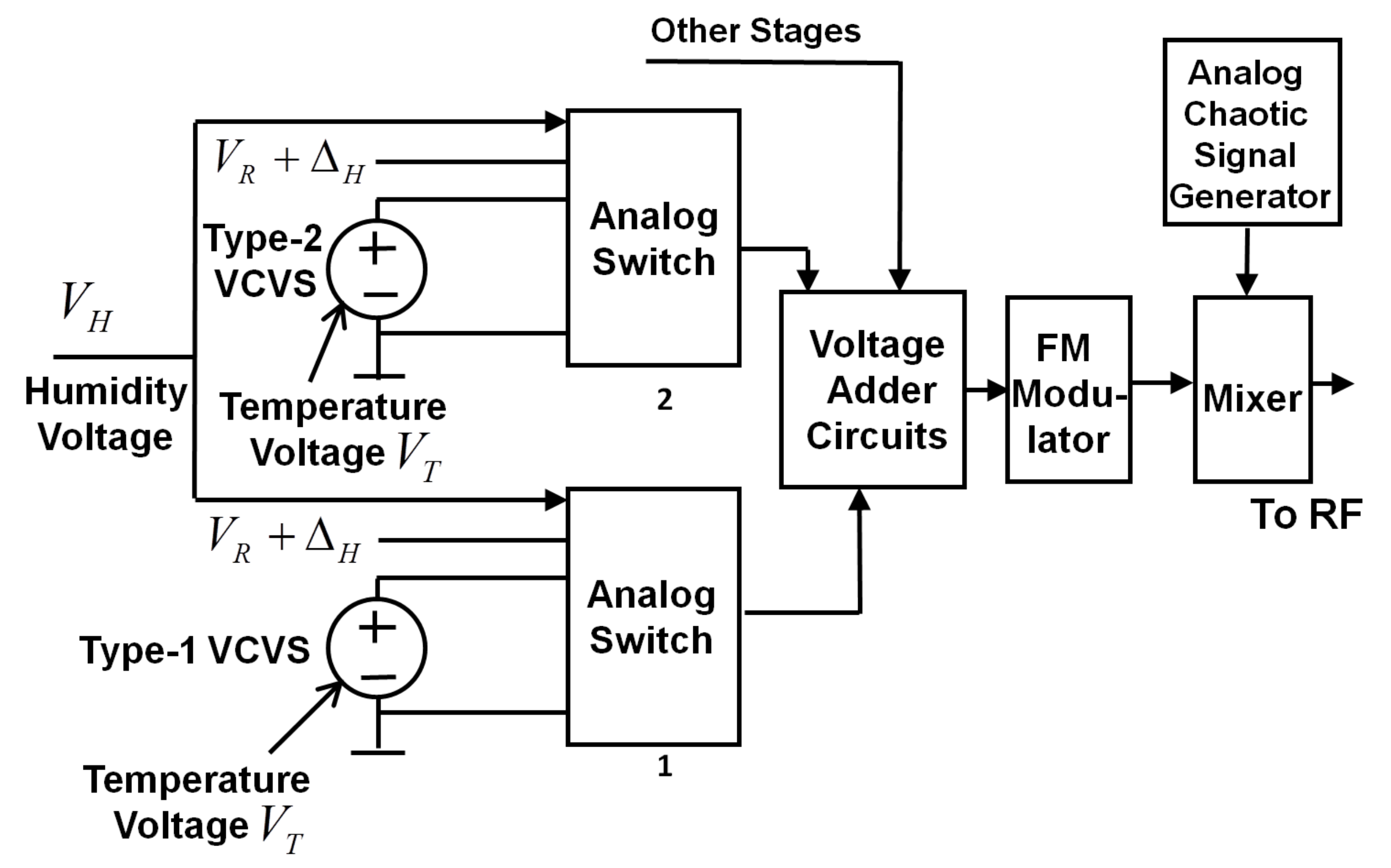}
\end{center}
\caption{Proposed Analog Circuit for Shannon's Rectangular Mapping (only one stage is shown). $V_H$ in comparison with threshold voltages generates select signals for the two analog multiplexers to decide which of the three inputs goes to the output. The outputs of both muxes are added to give this stage output. Similar outputs from higher stages are added to give AJSCC encoded voltage which is FM modulated and the mixed with semiorthogonal codes before RF transmission.}\label{fig:prop_ckt}
\end{figure}

With this information we can have the following circuit for each level to determine its contribution to the encoded length/voltage. Fig. \ref{fig:prop_ckt} shows the proposed circuit realization for level 1 and level 2 (stage 1). We can have an analog multiplexer that takes $0$, $V_R$ and $Type_{1,out}$ (for odd level)/$Type_{2,out}$ (for even level) as inputs and outputs one of these values based on it's select signals. Here $V_R$, the saturation voltage corresponds to the voltage that is proportional to the full length of the level. The select signals are generated by comparing $V_H$ with threshold voltages of the levels. Finally the voltage contributions from all levels are added to give the AJSCC encoded voltage which is then modulated by frequency position modulation, and sent to RF module for transmission.

It can be seen that our sensor node will be composed of low-cost mapping circuits, FM modulation circuits, and RF circuits. No microcontroller is needed, and no FPGA and DSP chips are needed either. Hence, the total power consumption of our design could be much lower than that of present digital sensor nodes, and the fabrication cost will also be much lower if fabricated on an Integrated Circuit (IC). A low-cost,
compact energy-harvesting unit for powering the sensor system can be used,
e.g., a tiny solar cell---given the sensor scale of a few
$\mathrm{cm^2}$---that can provide $\mathrm{mW}$-level power supply thus extending its lifetime to years. For more details about our prototype all-analog tier-1 sensor, including Spice, breadboard and Printed Circuit Board~(PCB) implementation and experimental results, please see \cite{Zhao16}.

\subsection{AJSCC Parameter Optimization}\label{sec:sim_ajscc}

\begin{figure}
\begin{center}
\includegraphics[width=3.4in]{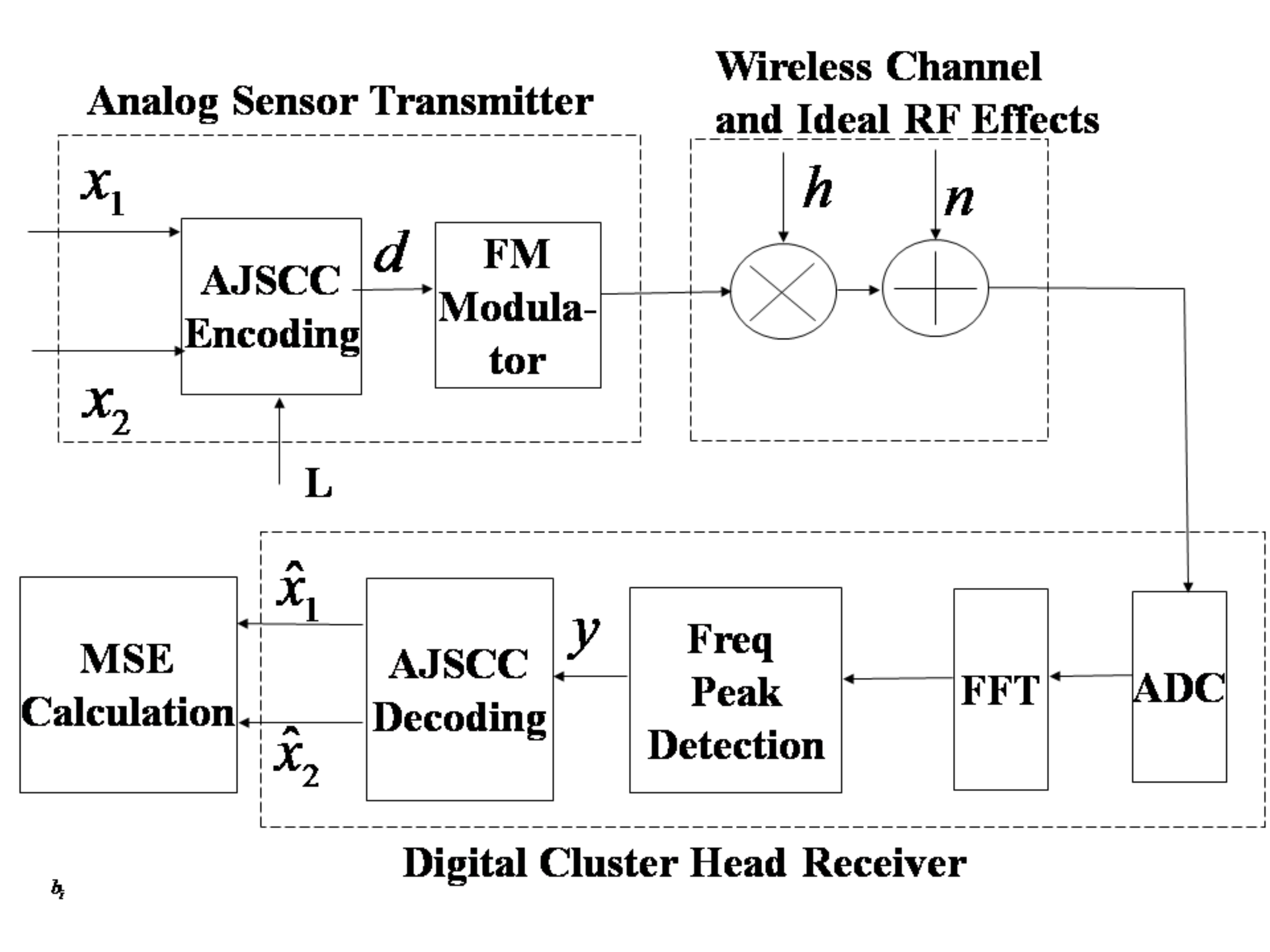}
\end{center}
\caption{Signal chain block diagram. AJSCC voltage is FM modulated and transmitted in the tier-1 sensor. Received baseband signal in tier-2 digital CH is sampled using ADC to find peak frequency using FFT. Peak frequency is mapped back to AJSCC voltage which is then decoded to sensor values.}\label{fig:system_paper}
\end{figure}

In this section, we theoretically analyze the AJSCC system to derive at optimum parameters (number of levels or the spacing $\Delta$ between parallel lines). Let us assume two independent sources, $X_1$ and $X_2$, whose distribution is unknown. The two sources are sensed and converted to voltages by the sensing units. Let's denote the two sensed signals by $x_1$ and $x_2$ and assume their ranges are $[0,V_{1}]$, and $[0,V_{2}]$ respectively. The AJSCC circuit with spacing $\Delta$ between parallel lines outputs the voltage $V_d$ given by,
\begin{equation}
V_d^{}  = \left\{ \begin{array}{l}
 kV_{1}^{}  + x_1^{} \,\quad \quad \quad \;k\;is\;even\, \\
 kV_{1}^{}  + V_{1}^{}  - x1^{} \quad k\;is\;odd \\
 \end{array} \right.
\end{equation}where the parameter $k = floor(\frac{x_2}{\Delta})$


We have an assumption that the encoded signal has an amplitude constraint $D_{max}$. This is because the FM modulator can only accept a signal within certain amplitude. If we denote the number of levels by $L$, this constraint can be denoted by $V_1^{} L \le D_{\max }^{}$. The extreme case will be $V_1^{} L = D_{\max }^{}$ where $V_1^{} L$ is the maximum output signal of AJSCC encoder. Due to this constraint, if we increase the parameter $L$, the voltage $V_1  = D_{max} / L$ for $x_1$ will be reduced. $V_2$ is a constant value, and the line spacing $\Delta = V_2 / (L-1)$.


The AJSCC encoding and decoding by the rectangular curve is depicted in Fig.~\ref{fig:system_paper}. The output of the AJSCC circuit is first frequency modulated and is then sent to RF circuitry for wireless transmission over a noisy wireless channel. At receiver, the baseband signal from RF is sampled by ADC and is sent to FFT block for frequency peak detection. Once the base-band frequency is determined, the AJSCC decoder finds the corresponding AJSCC voltage and then decodes it back to give reverse mapped signals $\hat x_1$ and $\hat x_2$.

\begin{figure}
\begin{center}
\includegraphics[width=3.4in]{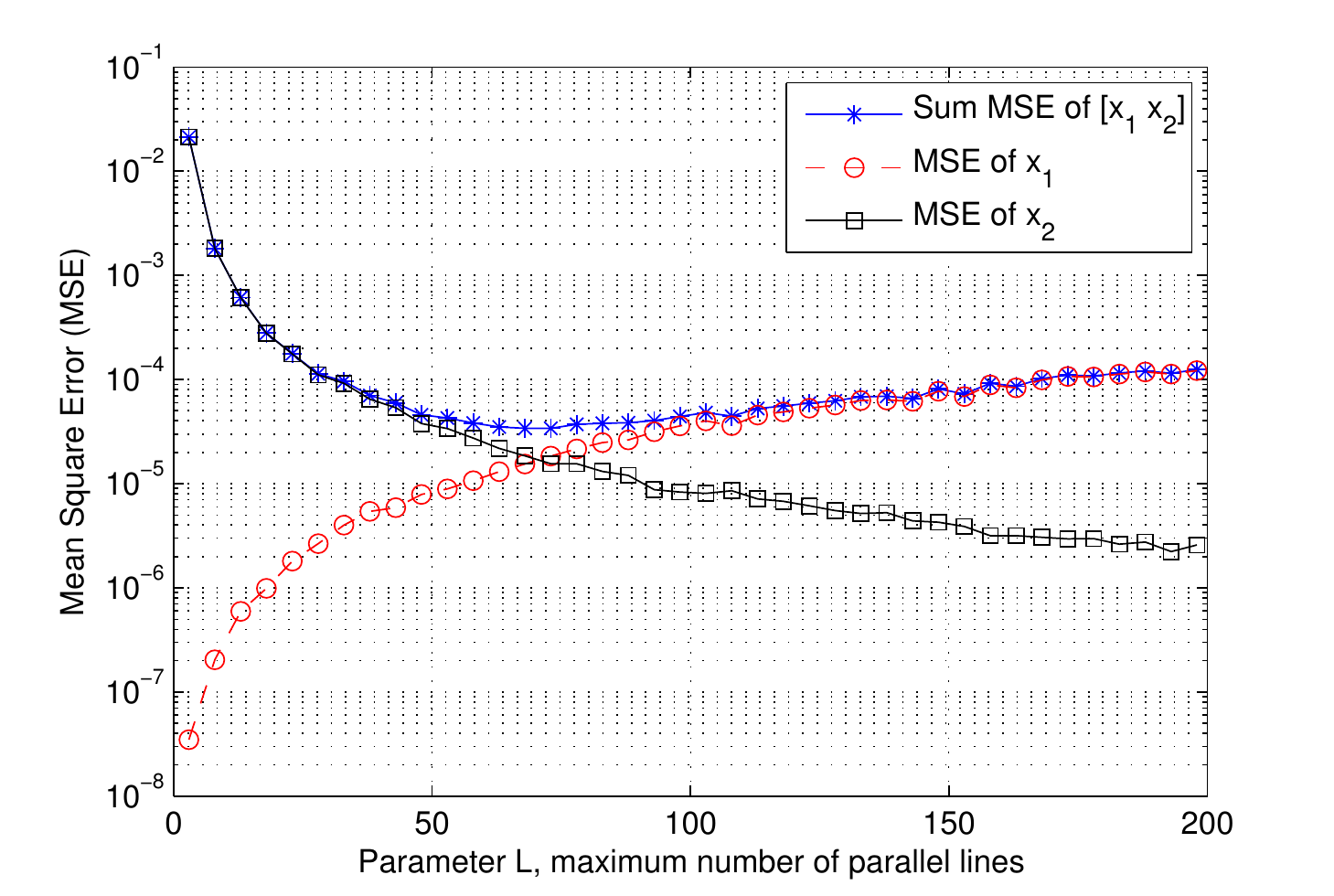}
\end{center}
\caption{MSE vs. parameter L, for SNR = $ -20 ~\mathrm{dB}$; Observed optimum L is about 73.}\label{fig:MSE_L}
\end{figure}

We have an interesting tradeoff behavior here. With increasing L, the MSE of $x_2$ drops as the spacing between the lines ($\Delta$) reduces. However, due to the constraint of the transmitted signal, the voltage representing $x_1$ will be smaller which will introduce higher error in MSE for $x_1$. With decreasing $L$, the quantization error in $x_2$ increases leading to high MSE for $x_2$d. This tradeoff behavior has been studied via MATLAB simulations to find an optimal $L$. We assumed two sources with uniform distribution between [0,1]. The parameter $D_{max}$ is set to 5. The signal $V_d$ generated by AJSCC mapping is FM modulated, in which a scaling factor is applied to convert voltage to frequency. Assuming linear transformation from voltage to frequency, and a scaling factor of 1000, the frequency range is from 0Hz to 5kHz. The SNR in the simulation is defined as the transmitted signal power to the noise power in the channel. The transmitted signal power is assumed to be unity for FM modulation of continuous cosine wave of amplitude equals to 1. The noise power is defined by the variance of the Gaussian noise. It is assumed that the channel is static channel between the sensor to the cluster head. Since there is no sensor movement and the environment is also static, it can be assumed that the channel gain is a constant value with phase shift. In the receiver, the ADC samples at a frequency of 65.536kHz and the frequency resolution of the signal is assumed as 1Hz giving FFT size as 65536. Once the peak frequency is determined, the AJSCC decoder decodes the baseband signal by first reverse mapping the frequency back to voltage and then voltage back to $\hat x_1$ and $\hat x_2$. Fig.~\ref{fig:MSE_L} shows the MSE-vs-L tradeoff behavior for an SNR value of -20 dB. We have observed that this FM modulated system achieves a low sum MSE of $3*10^{-4}$, but requires large number of parallel lines, around 73, to achieve this minimum MSE. The minimum sum MSE and the corresponding $L$ doesn't change much for SNRs of $-20dB$, $-10dB$ and $0dB$. The mapping can be extended to more than two sources. In three sources case, two sources will be discrete, and one will be continuous. Two modulus calculations need to be performed at the receiver for three source scenario. The above simulated system can be generalized to a sensor network, where different sensors transmit the FM modulated signal in overlapped frequency bands. The sensor signals are separated by semi-orthogonal signals mixed with the transmitted signal. In receiver, there will be interference from other sensors, thus the SNR will be SINR (Signal to Interference-plus-Noise Ratio). The level of interference is determined by the semi-orthogonal signals.

\begin{figure}
\begin{center}
\includegraphics[width=3.4in]{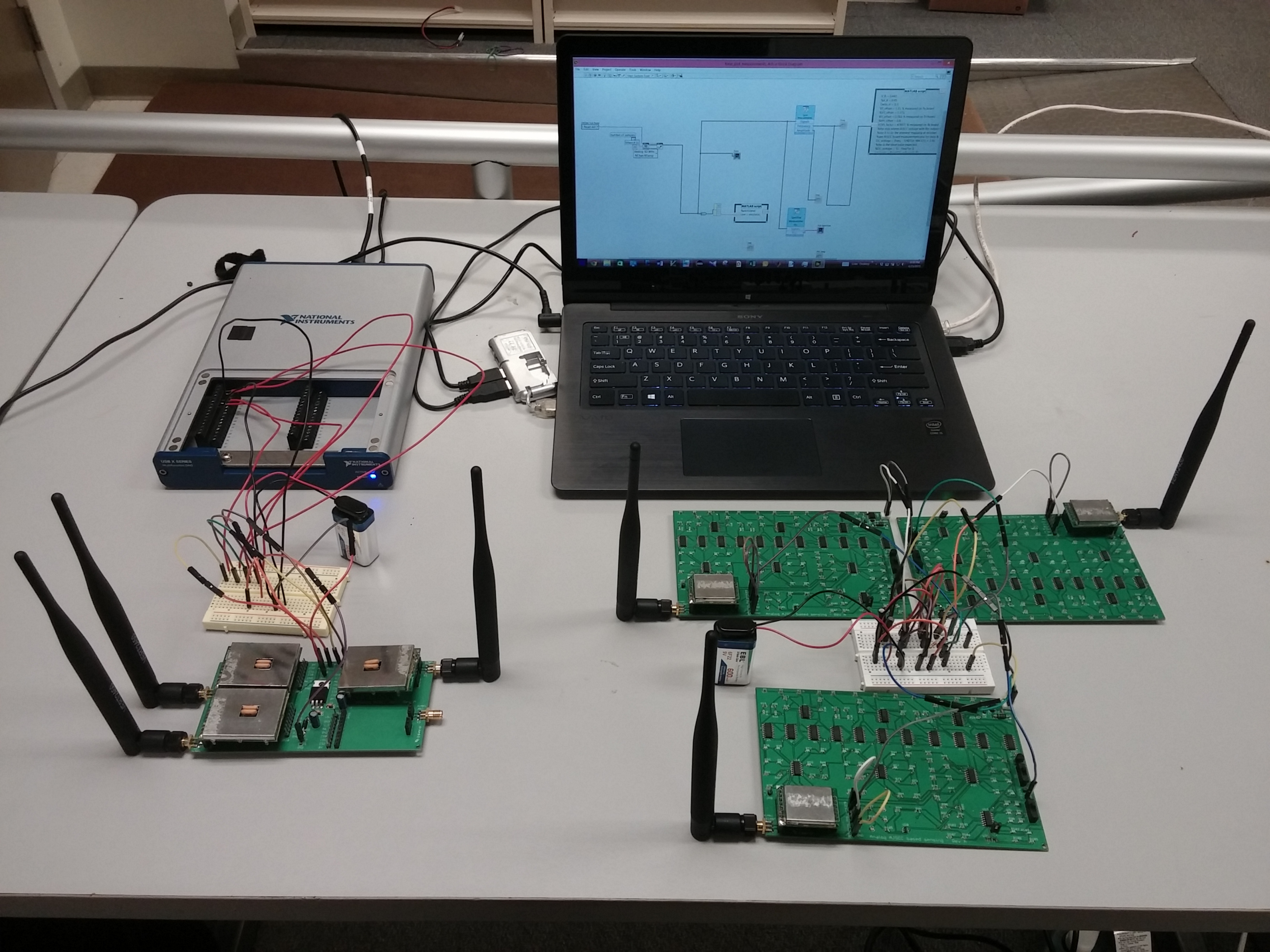}
\end{center}
\caption{Prototype: 3 Tier-1 sensors (right bottom) communicating to a Tier-2 receiver CH (left bottom). The baseband signals of all three channels are captured using NI Digital Acquisition System and processed/decoded on host computer using LabView/MATLAB.}\label{fig:prototype}
\end{figure}

\section{Performance Evaluation}\label{sec:perf_eval}

\begin{figure*}[ht!]
\centering
\begin{tabular}{ccc}
\hspace{-0.6cm}
\includegraphics[width=0.34\textwidth]{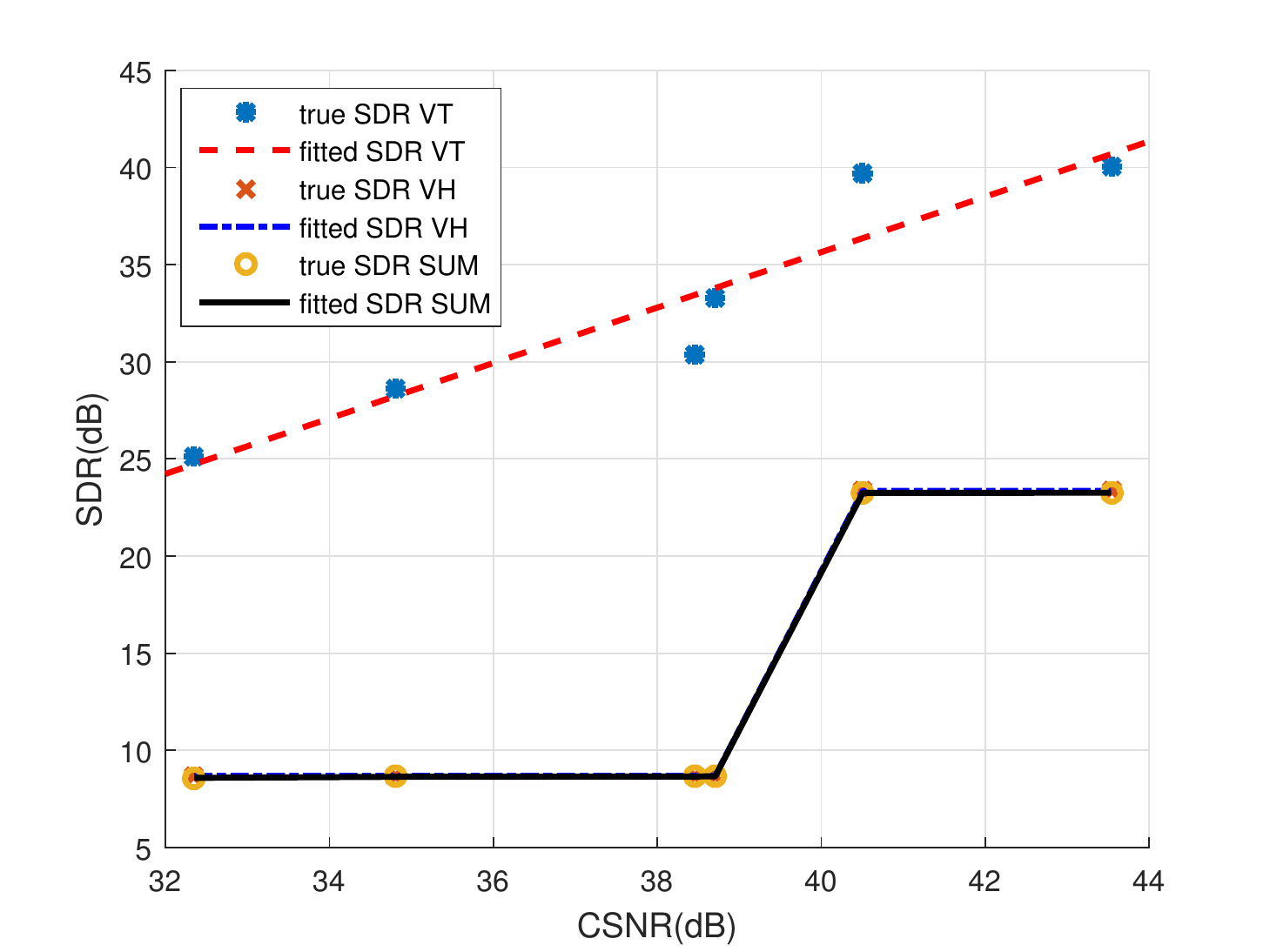} & \hspace{-0.5cm}
\includegraphics[width=0.34\textwidth]{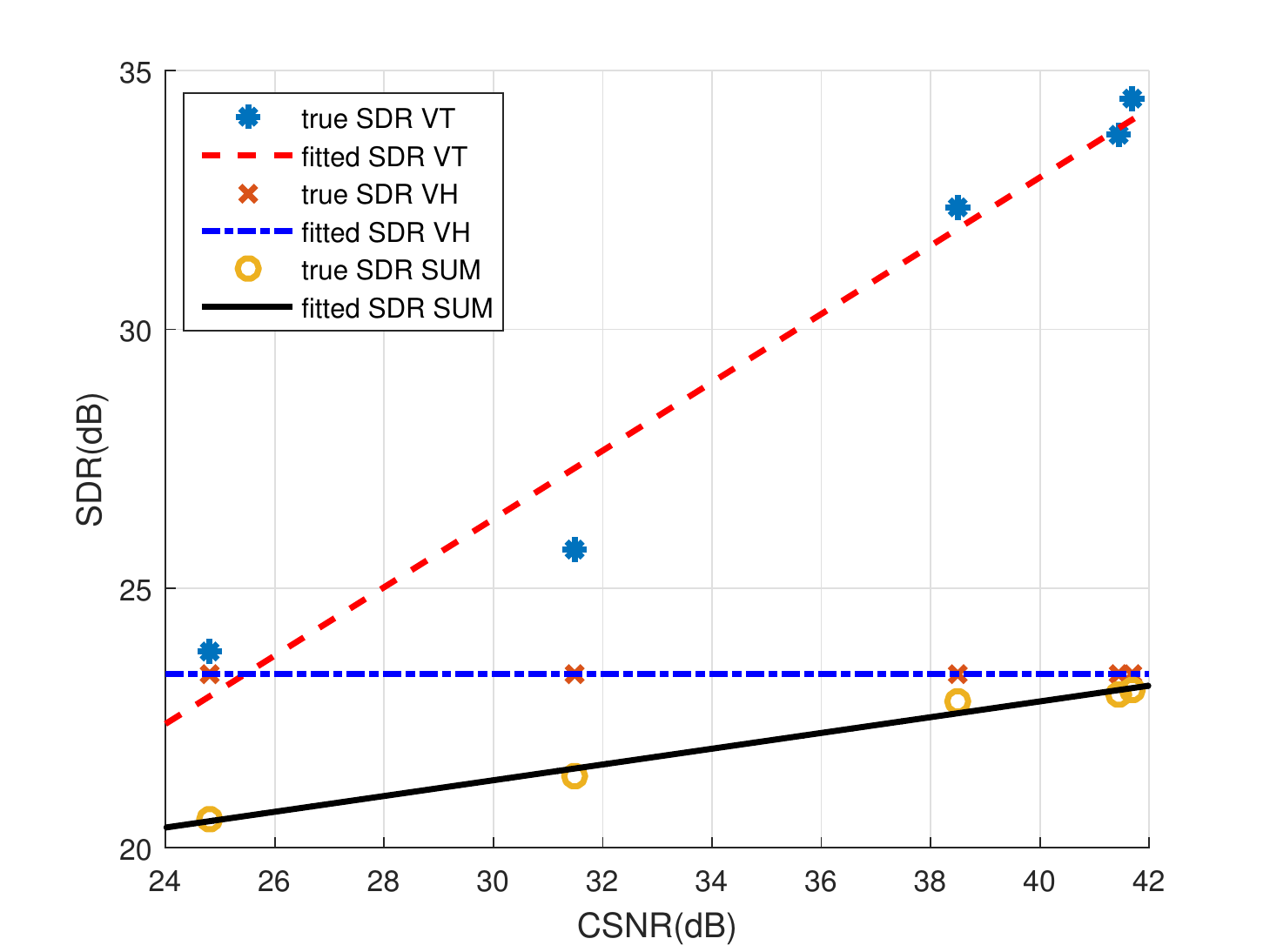} & \hspace{-0.5cm}
\includegraphics[width=0.34\textwidth]{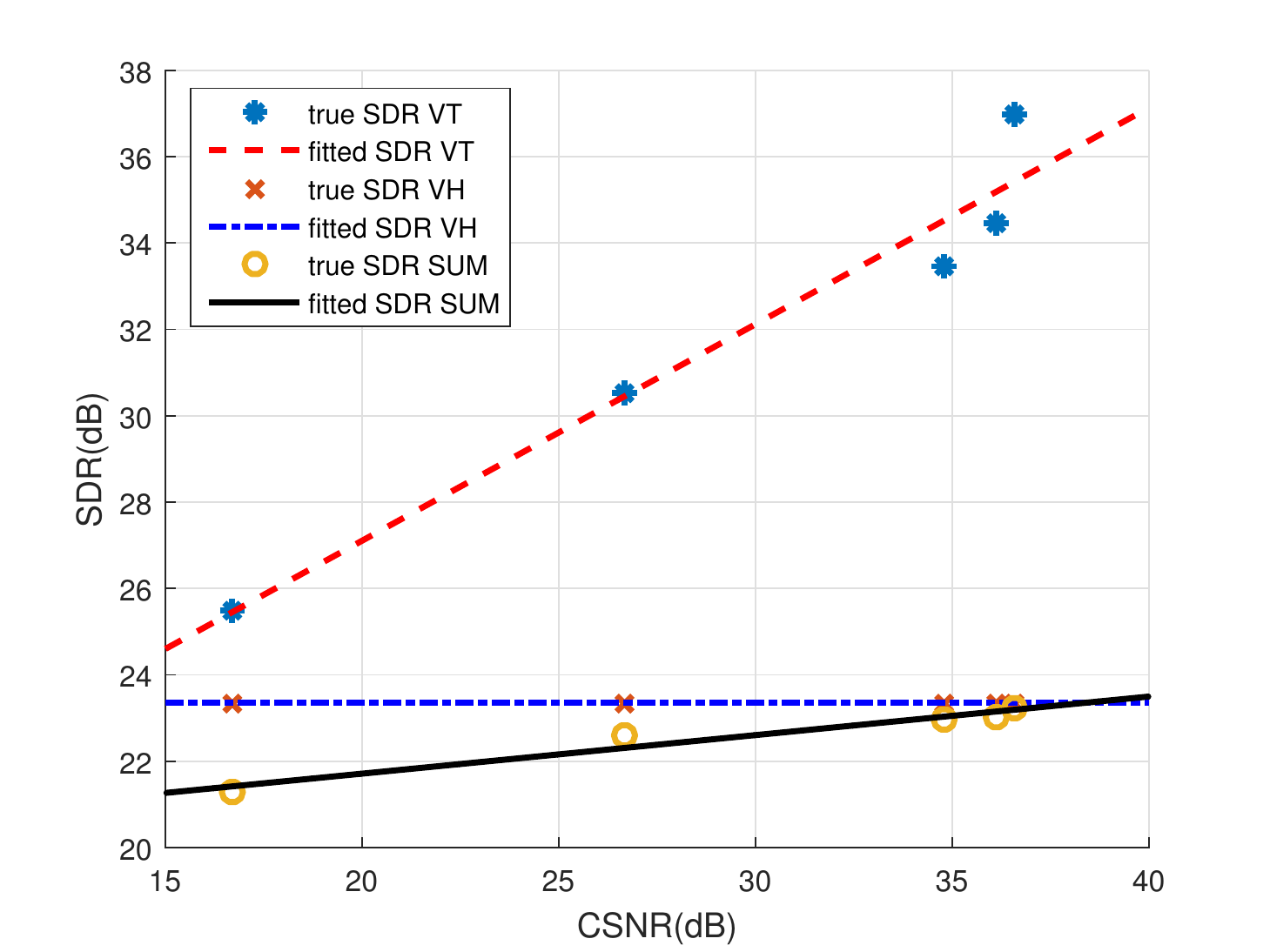}  \\
\hspace{-0.6cm}
\small (a)   & \hspace{-0.5cm} \small(b)    & \hspace{-0.5cm} \small(c)
\end{tabular}
\caption{SDR-vs-CSNR performance for different number of tier-1 sensors communicating to a digital cluster head using different channels: (a)1 sensor (b) 2 sensors (c) 3 sensors. Note the large jumps in SDR (seen in (a)) owing to similar behavior in SDR of $V_H$ which is because of the discreteness in $\hat V_H$}\label{fig:csnr_sdr_tx}
\end{figure*}

We first describe the Printed Circuit Board~(PCB) tier-1 sensor we developed along with power and cost analysis. Later we present some performance results of our Tier-1 sensor prototype when one, two and three of them communicate with a simple tier-2 receiver using FDMA. 

\begin{figure}
\begin{center}
\includegraphics[width=3.12in]{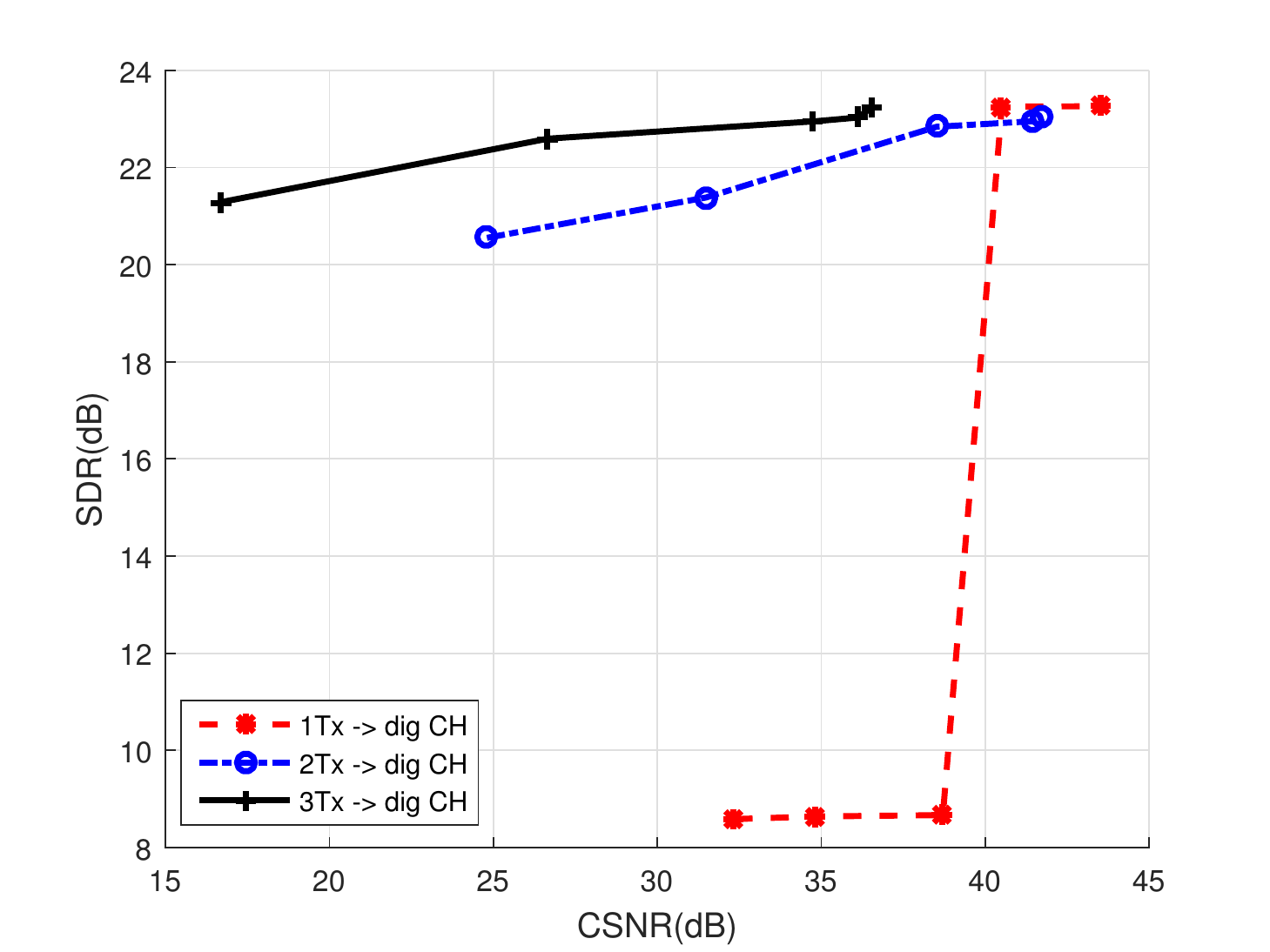}
\end{center}
\caption{Measured SDR-vs-CSNR performance for one, two and three Tier-1 sensors communicating with a Tier-2 CH. Due to receiver diversity, SDR of two and three sensor cases is high compared to one sensor case. Due to reduced performance, the one-sensor case exhibits a sharp decline in SDR value. The observed step is because the decoded $V_H$ values are discrete}\label{fig:csnr_sdr_combined}
\end{figure}

\textbf{Printed Circuit Board Sensor:}
We first tested our circuit idea (presented in Sect.~\ref{sec:prop_ckt}) on breadboard and obtained satisfactory results. This motivated us to go a step further to implement the full circuit (all stages) along with the RF part (as a COTS component) i.e., a full-fledged sensor on a PCB entirely designed by us (see Fig.~\ref{fig:prototype}). This tier-1 sensor consists of three major blocks - AJSCC encoding block, DC-to-sine wave conversion circuit and an RFIC module. The AJSCC encoding block takes temperature and humidity voltages as input and outputs the AJSCC encoded voltage. It implements 11 VCVS levels in total as per the setup described above ($\Delta_H=0.3V$). It also has option to take input from two external sources for testing and verification purposes. Then, a sine wave whose frequency is proportional to the AJSCC voltage is generated using a COTS timer chip. This sine wave is given as input to the COTS RF module which frequency modulates it and transmits the modulated signal at 2.4 GHz ISM band. FM requirement is to be noted here as we have shown in Sect. \ref{sec:sim_ajscc} that it has robust performance even at low SNR conditions.

\textit{Sensing Power and Cost Estimation:}
We analyze the power consumption of this tier-1 analog sensor by comparing it with that of existing sensors. State-of-the-art sensor nodes consume $\approx0.5~\mathrm{mA}$ in active mode and a few $\mathrm{\mu A}$ in sleep mode with supply voltages in the range $1.8-3.0~\mathrm{V}$, i.e., $0.9-1.5~\mathrm{mW}$ without taking into account the radio power: the active power consumption is mainly due to the microcontroller and Analog-to-Digital (A/D) conversions (\cite{wikiwsn},\cite{cosen}). We have listed some of the well-known wireless sensor nodes with their active MCU current consumption in Table.~\ref{table:compare} for comparison. In contrast, our all-analog sensor does not use power-hungry A/D's or microcontrollers (MCUs). The current drawn by the AJSCC baseband circuitry (using COTS OpAmps, Multiplexers, etc.), i.e., the entire board excluding the RFIC module, is $\approx3~\mathrm{mA}$ with a supply voltage of $5~\mathrm{V}$ (equivalent to $\approx15~\mathrm{mW}$); the cost of the AJSCC PCB is about $\$25$. These numbers, which are high because of (a) the use of COTS components and (b) duplication of hardware for each stage, can be reduced \emph{drastically} if Integrated Circuit~(IC) design is adopted. While our implementation serves as feasibility study, we believe the power consumption can be reduced to $<50~\mathrm{\mu W}$ if (1) our circuit is redesigned using the latest nm-technology components (for OpAmps, Comparators, and Multiplexers) (2) our circuit is redesigned integrating all the functionality into a single Integrated Circuit (IC) rather than using discrete COTS components (3) hardware duplication issue is solved and we have some preliminary ideas too on that front (possibility of $<10~\mathrm{\mu W}$)

Let us provide a rough estimate using just (1) above: our circuit in total (5 and half stages/11 levels) consists of 16 OpAmps, 17 Comparators, and 11 Multiplexers, where OpAmps are clearly the major contributors to the overall power consumption. There are many low-power designs proposed for these components. For example, a low-power OpAmp~\cite{opamp} consuming about $8~\mathrm{\mu W}$, a comparator~\cite{comparator} consuming about $12.7~\mathrm{nW}$, and an analog multiplexer ($ADG704$) consuming about $10~\mathrm{nW}$ can be used for our circuit resulting in a power consumption of $\approx\mathrm{130 \mu W}$. We are also optimistic that the sensor cost would reduce to less than $\$5$ leveraging economies of scale via mass production using the latest IC technology. Achieving both goals will enable critical futuristic applications such as persistent wireless sensing and IoT-based solutions.
%
 %
%
%

\textbf{Prototype Performance:}
We have developed a simple prototype (Fig.~\ref{fig:prototype}) consisting of three tier-1 sensors communicating to a simple tier-2 Cluster Head (CH). 
The receiver CH, also designed by us, has three antennas for receiving the signals of the three sensors. There are three RF receivers to downconvert the RF signals received by the three antennas. The baseband signals at the output of RF receivers (which are supposed to be sine waves as in the case of Tx) are then fed to NI Digital Acquisition (DAQ) system to detect their peak frequency in a LabView program. MATLAB is used inside LabView to perform spectral analysis of the signal (such as SNR calculation) and also to map back the detected peak frequency to DC voltage (AJSCC voltage) and then decode the AJSCC voltage back to temperature and humidity voltages, [$\hat{V_T},\hat{V_H}$]. Mean Square Error (MSE) and the Signal to Distortion Ratio (SDR) has been calculated as follows.
\begin{align*}
  MSE & = (V_T - \hat V_T)^2 + (V_H - \hat V_H)^2 \\
  SDR & = 10\log_{10}(\frac{1}{MSE})
\end{align*}
%
%
%
%
%

We measured and compared the prototype's performance for cases when one, two and three sensors are communicating simultaneously to digital CH using FDMA (different channels). Fig.~\ref{fig:csnr_sdr_tx}(a), Fig.~\ref{fig:csnr_sdr_tx}(b), Fig.~\ref{fig:csnr_sdr_tx}(c) show the SDR-vs-CSNR performance for these three cases respectively. Here CSNR (Channel Signal-to-Noise Ratio) is the SNR of the baseband signal at the output of receiver RF module. SDR and CSNR values are plotted by varying the distance between Tx and Rx and fixing $V_T$ and $V_H$. 
We observe that at high CSNR, combined SDR is limited by that of $V_H$. This is because $V_H$ is near the threshold voltage (rather than level voltage) resulting in large quantization error in $V_H$ and so very less SDR. Also in (a), we see a step, this is because SDR of $V_H$ is discrete. When CSNR varies, $V_H$ is decoded to discrete levels instead of a continuous value resulting in discrete variation in its SDR. The decoded $V_H$ value spreads over two levels in (a) (showing a huge step) while it is at single level in case of (b) and (c). 

Figure~\ref{fig:csnr_sdr_combined} compares the SDR-vs-CSNR performance of one, two and three sensors communicating to digital CH using different channels showing the effect of receiver diversity. Sum SDR for the three cases in Fig.~\ref{fig:csnr_sdr_tx} are plotted in a single figure for comparison. We can clearly observe that, three sensor case has better performance than 2 sensor case which in turn is far better than one sensor case due to receiver diversity. Also SDR of one sensor case quickly diminishes as SNR is reduced (due to discreteness in $V_H$) while the other two cases are relatively robust against this behavior. Since our architecture allows high density deployment, we believe the benefits of receiver diversity can be harvested. Finally these results show that (i) it is indeed possible to build a low-power and low-cost tier-1 sensor (ii) several such tier-1 sensors can communicate to a tier-2 CH using FDMA. We would like to mention that we designed tier-1 analog sensors only as a \textit{feasibility study} (we are not electrical engineers to design it perfectly). We are confident that far impressive results can be achieved with dedicated IC design for tier-1 sensors that would also significantly reduce cost and power.

\section{Conclusions and Future Work}\label{sec:conc}
A novel multi-tiered architecture has been proposed for Wireless Sensor Networks~(WSNs) that separates sensing and computational aspects. In order to achieve low-power and low-cost objectives, a sensing paradigm that is based on AJSCC (Shannon Mapping) has been used for Tier 1 whose main function is to sense, encode and transmit values (dumb sensing with no intelligence) to Tier 2 consisting of resource-rich digital cluster heads with powerful signal processing capabilities. 
We have also proposed a simple analog circuit to realize the rectangular type of AJSCC mapping. This circuit for tier-1 sensors has also been realized on a Printed Circuit Board for feasibility study.
A simple prototype consisting of three these Tier-1 sensors communicating to a simple Tier-2 receiver using FDMA has been demonstrated to satisfactorily prove the feasibility of our \textit{low-power, all-analog sensing} idea. 

As future work, we will develop ``smart processing'' algorithms at tier-2 for fault-detection, denoising, filtering, etc. We will also investigate the use of this framework to monitor a full-scale bridge superstructure subjected to accelerated aging at Rutgers University. We will carry this out at a unique facility, the Bridge Evaluation and Accelerated Structural Testing~(BEAST)~\cite{BEAST}, constructed by the Center for Advanced Infrastructure and Transportation~(CAIT). Tier-1 sensors in this case, will be installed at various places on the bridge to sense and transmit pressure/strain data to mobile Tier-2 CHs which will process this data to extract meaningful information. We also plan to further investigate using Frequency Position Modulation~(FPM) for Tier-1 sensor multiplexing and also realize it in hardware to assess the true performance. 


\balance

\bibliographystyle{abbrv}
\bibliography{MyPublications_v1,career_v1,sensor_v5,references_v3.0,cross-layer_v3,underwater_v19,RobustConsensus}

\end{document}